\documentclass[aps,pre,amsmath,showpacs]{revtex4}
\usepackage{graphicx}% Include figure files

\usepackage{amsmath}
\usepackage{amssymb}
\usepackage{graphicx}

\newcommand{\eps}{\varepsilon}
\newcommand{\pd}{\partial}
\newcommand{\bx}{{\bf x}}

\newcommand{\be}{\begin{equation}}
\newcommand{\ee}{\end{equation}}
\newcommand{\Fa}{\vphantom{\vdots}}
\newcommand{\berr}{\be\begin{array}{c}}
\newcommand{\berrl}{\be\begin{array}{l}}
\newcommand{\eerr}{\Fa\end{array}\ee}
\newcommand{\eerrl}{\end{array}\ee}
\def\eps{\varepsilon}

\def\k{{\bf k}}
\def\x{{\bf x}}

\def\p{{\bf p}}

\newcommand{\beq}{\begin{eqnarray}}
\newcommand{\eeq}{\end{eqnarray}}

\begin{document}

\title{{Two-loop calculation of the turbulent Prandtl number}}
\author{
L. Ts. Adzhemyan$^1$, J. Honkonen$^2$, T. L. Kim$^1$, and L. Sladkoff$^1$}

\affiliation{$^1$ Department of Theoretical Physics, St.~Petersburg
University, Uljanovskaja 1, St.~Petersburg,
Petrodvorets, 198504 Russia, $^2$Department of Technology, National Defence
College, P.O.~Box~7, FI-00861, Helsinki and
Theoretical Physics~Division,
Department~of~Physical Sciences, P.O.~Box~64, FI-00014
University~of~Helsinki, Finland}

\begin{abstract}
The turbulent Prandtl number has been calculated in the two-loop approximation
of the $\eps$ expansion of the stochastic theory of turbulence. The strikingly
small value obtained for the two-loop correction explains the good agreement of the earlier
one-loop result with the experiment. This situation is drastically different from
other available nontrivial two-loop results, which exhibit corrections of the magnitude of the
one-loop term. The reason is traced to the mutual cancellation of additional divergences
appearing in two dimensions which have had a major effect on the results of previous calculations
of other quantities.
\end{abstract}

\pacs{%47.27.$-$i,%Turbulent flows, convection, and heat transfer
47.27.Te, %Convection and heat transfer
05.20.Jj, %Statistical mechanics of classical fluids
05.10.Cc %Renormalization group methods
}

\date{\today}

\maketitle

\section{\label{sec:intro}Introduction}

The method of renormalization group (RG) in the theory of
developed turbulence is currently the most developed technical
means allowing for reorganization of the straightforward
perturbation theory, whose huge expansion parameter at large
Reynolds numbers renders it practically useless. At the same time
the physical value of the artificial expansion parameter $\eps$
introduced in the RG approach is not small either. For some
important physical quantities, such as the critical dimension of
the velocity and effective viscosity, it is possible to prove with
the use of of Galilei invariance of the theory that the
corresponding series in $\eps$ terminate at the linear terms.
Therefore, for such quantities the RG approach yields exact
answers coinciding with the prediction of the phenomenological
theory of Kolmogorov. For other interesting quantities, such as
the Kolmogorov constant, skewness factor, turbulent Prandtl number
and the like, the series in $\eps$, however, do not terminate. In
this context, it has been often suggested that with the aid of the
$\eps$ expansions it is not possible to obtain a sufficiently good
estimates of numerical values of these quantities, although --
until recently -- there were no calculations extending beyond the
first order of the perturbation theory (one-loop approximation).
The two-loop calculation of the Kolmogorov constant and the
skewness factor in the inertial range carried out in Ref.
\cite{2loop} confirmed this pessimistic point of view on the
whole: the two-loop contribution turned out to be practically
equal to the one-loop contribution, although the trend of change
of the quantities calculated was correct, i.e. towards the
experimental value from the one-loop result.

In Ref. \cite{2loop} calculations were carried out for space dimensions $d$ different from
$d=3$ as well. It turned out that the relative magnitude of the two-loop contribution decreases with the
growth of $d$, and in the limit $d\rightarrow \infty$ is of the order 10 \% only. At the same time
in the limit $d\rightarrow 2$ this contribution grows without limit. Such a behavior of the coefficients of the
$\eps$ expansion may be related to that their singularities as functions
$d$ lie in the region $d \leq 2$. The nearest singularity at $d=2$ is connected with the
divergence of some graphs in the limit $d \rightarrow 2$, which leads to the appearance of poles in
$d-2$ in the coefficients of the $\eps$ expansion, and it is just these graphs which turn out to be
responsible for the large value of the two-loop contribution at $d=3$. This feature gave rise to the hope that
summation of these singularities may lead to quantitative improvement of the results of the $\eps$ expansion in
the real dimension $d=3$. Such a summation was successfully carried out in the framework of the RG method with
the aid of the account of the additional UV renormalization of the theory in the vicinity of $d=2$ \cite{improved}.
In the resulting ''improved $\eps$ expansion'' the low-order terms are calculated in the usual way at $d=3$, while the
high-order terms are approximately summed with the account of their leading singularities in
$d-2$ (one-loop approximation), then next-to-leading singularities (two-loop approximation) etc. Calculation of
the Kolmogorov constant and skewness factor according to this program has demonstrated an essential decrease
of the relative impact of the two-loop contribution and led to a fairly good agreement with the experiment
\cite{improved}.

In the present paper we shall analyze to what extent the singularities of the
$\eps$ expansion show for another important characteristic quantity of turbulent systems, the
turbulent Prandtl number. It was calculated in the framework of the RG and the
$\eps$ expansion in Refs. \cite{Fournier,Hnatich} with rather good agreement with the experiment
\cite{Monin,Chua90,Chang02}. We have carried out a two-loop calculation of the Prandtl number in order to check, whether this
agreement is partially coincidental.

Let us remind that the Prandtl number is the dimensionless ratio of the coefficient of kinematic viscosity
$\nu_0$ to the coefficient of thermal diffusivity ${\cal \kappa}_0$. (In the formally identical problem of
turbulent diffusion the ratio of the coefficients of kinematic viscosity and diffusion is called Schmidt number).
For systems with strongly developed
turbulence the process of homogenization of the temperature is strongly accelerated,
which is reflected in the value of the effective
or turbulent coefficient of thermal diffusivity. The ratio of the coefficient of turbulent viscosity and the coefficient of
turbulent thermal diffusivity is the turbulent Prandtl number. Contrary to its molecular analog the turbulent Prandtl number is
universal, i.e. does not depend on individual properties of the fluid. For the accurate determination of the turbulent
Prandtl number a set of conditions is required, especially when calculations are carried out in the two-loop
approximation. Therefore, apart from the formulation of the stochastic problem we shall
pay the proper attention to this problem as well.

The present paper is organized as follows. In Sec. \ref{sec:model} we remind the main features of the
description of passive advection of a scalar quantity in the stochastic theory of turbulence with special
emphasis on the careful definition of the turbulent Prandtl number within the model considered. Sec. \ref{sec:R}
is devoted to the analysis of renormalization and renormalization-group equations of the model. In Sec. \ref{sec:Pr}
details of the two-loop calculation are presented. Sec. \ref{sec:Zakl} contains analysis of the results and
concluding remarks.

\section{\label{sec:model}Description of the model}

Turbulent mixing of a passive scalar quantity is described by the equation
\begin{equation}\label{1}
  \pd_t \psi+(\varphi_j\pd_j)\psi= {\cal \kappa}_0 \Delta  \psi + f \,.
\end{equation}
The field $\psi(\x,t)$ in Eq. (\ref{1}) may have the meaning of both the non-uniform temperature
($\kappa_0$ being the coefficient of thermal diffusivity)
and concentration of the particles of the admixture (in this case
${\cal \kappa}_0$ is replaced by the coefficient of diffusion). The field $f(\x,t)$ is the source of the passive scalar field.
In the stochastic model of turbulence the field of turbulent eddies of the velocity of the incompressible fluid
$\varphi_i(\x,t)$ satisfies the Navier-Stokes equation with a random force:
\begin{equation}\label{2}
  \pd_t\varphi_i+(\varphi_j\pd_j)\varphi_i=\nu_0\Delta \varphi_i-\pd_i
  P+F_i \,,
\end{equation}
where $P(t,\bx)$ and $F_i(t,\bx)$ are, respectively, the pressure and the transverse external random
force per unit mass. For $F$ Gaussian distribution with zero mean and the correlation function
\begin{equation}
\langle F_i(t,\bx)F_j(t',\bx')\rangle =\delta (t-t')(2\pi )^{-d}\int d{\bf
k}\, P_{ij}({\bf k}) d_F({k})\exp [i{\bf k}({\bf x}-{\bf x'})]
\label{3}
\end{equation}
is assumed. Here, $P_{ij}({\bf k}) =\delta _{ij}  - k_i k_j / k^2$ is the transverse projection operator,
$d_F(k)$ a function of $k\equiv |{\bf
k}|$ and parameters of the model, and $d$ the dimension of the coordinate space ${\bf x}$.

The stochastic problem (\ref{1}) -- (\ref{3}) is equivalent to the quantum-field model
with the doubled number of fields $\phi\equiv\{\varphi,\psi, \varphi',
\psi'\}$ and the action
\begin{equation}
S(\Phi )=\varphi 'D_F \,\varphi '/2+\varphi '[-\partial _t\varphi
+\nu _0 \Delta  \varphi -(\varphi \partial )\varphi ] + \psi'
[-\partial _t\psi +{\cal \kappa}_0 \Delta  \psi -(\varphi
\partial )\psi + f]  , \label{action}
\end{equation}
in which $D_F$ is the correlation function of the random force (\ref{3}) and
the necessary integrations over $\{t, {\bf x}\}$ and summations over vector indices are implied.
In model (\ref{1})--(\ref{action}) only correlation functions of the admixture field of the form
\be \langle
\psi(\x_1,t_1),\psi(\x_2,t_2)...\psi(\x_n,t_n) \psi
'(\x'_1,t'_1),\psi '(\x'_2,t'_2)...\psi '(\x'_n,t'_n)\rangle\, ,
\nonumber \ee
with the meaning of multiple response functions are nonvanishing. The simplest of them is determined
by the following variational derivative with respect to the source
$f$ in Eq. (\ref{1})
\be
G(\x-\x',t-t') \equiv \langle \psi(\x,t) \psi '(\x',t')\rangle\big |_{f=0}
= \frac {\delta \langle\psi(\x,t)\rangle}{\delta f(\x',t')}\bigg |_{f=0}\,.
\label{Gxt}
\ee
The non-random source field $f$ of the passive scalar has been introduced in action
(\ref{action}) solely to remind of relation
(\ref{Gxt}) and its generalizations and will therefore further be omitted.

Model (\ref{action}) gives rise to the standard diagrammatic
technique with the following nonvanishing bare propagators
($t\equiv t_1-t_2$)
\begin{align}
\langle \varphi(t_1) \varphi(t_2)
\rangle _0&={d_F(k) \over 2\nu_0
k^{2}}\,\exp \left(-\nu_0 k^{2} |t| \right) ,\label{lines0}\\
 \langle \varphi(t_1) \varphi '(t_2)\rangle _0&=
\theta(t) \exp \left(-\nu_0 k^{2} t \right),
\label{lines}\\
\langle \psi(t_1) \psi '(t_2)\rangle _0&= \theta(t) \exp
\left(-{\cal \kappa}_0 k^{2}t\right), \label{lines1}
\end{align}
in the ($t$, ${\bf k}$) representation. The common factor
$P_{ij}({\bf k})$ has been omitted in expressions (\ref{lines0}) --  (\ref{lines}) for simplicity. Interaction in
action (\ref{action}) corresponds to the three-point vertices \,
$-\varphi'(\varphi\partial)\varphi=\varphi'_iV_{ijs}\varphi_j\varphi
_s/2$ \,
with the vertex factor \,$V_{ijs}={\rm i}(k_j\delta _{is}+k_s\delta
 _{ij})$,\, and \,$-\psi'(\varphi\partial)\psi={\rm i} k_j \psi' \varphi_j
 \psi$,\,
where ${\bf k}$ is the wave vector of fields $\varphi'$ and $\psi'$.

Turbulent processes lead to significantly faster than in relations
(\ref{lines}) and (\ref{lines1}) attenuation in time of the response functions
$\langle \varphi \varphi '\rangle$ and $\langle \psi \psi '\rangle$ due to
the effective replacement of the molecular coefficients of viscosity and thermal diffusivity by their
turbulent analogs. At the same time, however, the simple exponential time-dependence is changed
as well (and in a different manner for
$\langle \varphi \varphi '\rangle$ and $\langle \psi
\psi '\rangle$), therefore it is necessary to choose a definite way of fixing the ratio of the
turbulent transport coefficients, i.e. the Prandtl number (or Schmidt number). Henceforth, we shall use the following
definition. Consider the Dyson equations for the response functions in the wave-vector-frequency representation:
\begin{align}
G^{-1}_{\varphi \varphi'}(k,\omega) \equiv
\Gamma_{\varphi\varphi'}(k,\omega) &= - {\rm i} \omega + \nu_0 k^2
- \Sigma_{\varphi'\varphi}(k,\omega)\,, \label{Gff}\\
G^{-1}_{\psi\psi'}(k,\omega) \equiv \Gamma_{\psi\psi'}(k,\omega) &=
- {\rm i} \omega + {\cal \kappa}_0 k^2 -
\Sigma_{\psi'\psi}(k,\omega)\,, \label{Gtt}
\end{align}
where $\Sigma$ are the corresponding self-energy operators, and introduce the inverse effective
Prandtl number $u_{eff}$ by the relation
\begin{eqnarray}
u_{eff} \equiv \frac
{\Gamma_{\psi\psi'}(k,\omega=0)}{\Gamma_{\varphi\varphi'}(k,\omega=0)}
 \label{ueff}\,.
\end{eqnarray}
Further, we shall be interested
in the inertial range $L^{-1}\ll k\ll \Lambda $ (here, $L$ is the external scale of turbulence and
$\Lambda^{-1}$ the characteristic length of the dissipating eddies)
in which the quantity $u_{eff}$ is independent of the wave number $k$.

\section{\label{sec:R}Renormalization of the model and the RG representation}

The self-energy operators
 $\Sigma_{\varphi ', \varphi}(k,\omega)$ and $\Sigma_{\psi'\psi}(k,\omega)$
appearing in Eqs. (\ref {Gff}) and (\ref {Gtt}) may be found in model
(\ref{action}) in perturbation theory. However, the expansion parameter turns out to
be very large for developed turbulence (for $\Lambda L \gg 1$). The renormalization-group method
allows to carry out a resummation in the straightforward perturbation theory. To apply it, it is necessary
to use in relation (\ref{3}) ''the pumping function'' $d_F({k})$ of a special form
\be
d_F(k)=D_0 k^{4-d-2\eps}. \label{nakach}
\ee
In the infrared region the power function (\ref{nakach}) is assumed to be cut off
at wave numbers $k\leq m \equiv L^{-1}$. The quantity $\eps > 0$ in Eq.
(\ref{nakach}) is the formal small expansion parameter in the RG approach with the
value $\eps=2$ corresponding to the physical model.

The usual perturbation theory is a series in powers of the charge
$g_0\equiv D_0/\nu _0^3$ dimensionless at $\eps=0$ (logarithmic theory). At $\eps\rightarrow 0$
ultraviolet divergences are brought about in the graphs of the perturbation theory which show
in the form of poles in $\eps$. Due to Galilei invariance of the model divergences at
$d > 2$ are present only in the one-irreducible functions
$\langle \varphi \varphi '\rangle$ and  $\langle \psi \psi
'\rangle$  and are of the form $\varphi'\Delta \varphi$ and
$\psi'\Delta \psi$. At $d = 2$ also the one-irreducible function
$\langle \varphi' \varphi '\rangle$ diverges. For $d > 2$ the renormalized action may be written as
\[
S_R(\Phi )={1\over 2}\,\varphi 'D_F \varphi '+\varphi
'[-\partial _t\varphi +\nu Z_\nu \Delta  \varphi -(\varphi \partial
)\varphi ] + \psi' [-\partial _t\psi +u\, \nu Z_1 \Delta  \psi
-(\varphi
\partial )\psi ]\,.
\]
It is obtained from action (\ref{action}) by the multiplicative renormalization of the parameters of the
model:
\begin{equation}
\nu_0=\nu Z_{\nu}, \quad g_{0}=g\mu^{2\eps}Z_{g},\quad
u_0=uZ_u,\quad Z_u = Z_1 Z_{\nu}^{-1}, \quad Z_{g}=Z_{\nu}^{-3}
 \label{Z}
\end{equation}
with two independent renormalization constants $Z_{\nu}$ and $Z_{1}$.
The quantities $\nu$ and $g$ in Eq. ($\ref{Z}$) are the renormalized analogs of the
coefficient of viscosity and the coupling constant (the charge $g$ being dimensionless).
The renormalization mass $\mu$ is an arbitrary parameter of the renormalized theory, and
the pumping function $d_F(k)$ (\ref{nakach}) determining the correlation function of the random force
$D_F$ (\ref{3}) is assumed to be expressed in terms of the renormalized parameters:
\[
d_F(k)=g_0 \nu_0^3 k^{4-d-2\eps}= g \mu^{2\eps} \nu ^3
k^{4-d-2\eps}.
\]
The dissipative wave number $\Lambda$ is determined by
$g_0$ according to the relation $\Lambda = g_0^{1/2\eps}$. It may be also estimated by the quantity
$\mu$. Thus, the inertial range we are interested in corresponds to the condition $s \equiv k/\mu \ll 1$.

In the scheme of minimal subtractions (MS) used in the following the renormalization constants
have the form of the Laurent expansion $1 + \text{poles in } \eps$
\begin{eqnarray}
Z=1+\sum _{k=1}^{\infty }a_k(g,u)\eps ^{-k}=1+\sum _{n=1}^{\infty
}g^n \sum _{k=1}^{n}a_{nk}(u)\eps ^{-k}\,. \label{ZZ}
\end{eqnarray}
For $Z_{\nu}$ at $d=3$ in Ref. \cite{Pismak} the following expression was obtained
\begin{eqnarray}
Z_{\nu}=1 + \frac{a_{11}^{(\nu)} g}{\eps} + O(g^2)\,, \qquad
a_{11}^{(\nu)}= - \frac{(d-1)\bar S_{d}}{8(d+2)}\, ,\quad \bar
S_{d}\equiv {S_d\over (2\pi)^d}\,,\label{Znu}
\end{eqnarray}
where  $S_d=2\pi^{d/2}/\Gamma(d/2)$  is the area of the $d$-dimensional sphere of unit radius.

The correlation functions of the renormalized theory do not contain poles in
$\eps$. This feature, however, does not solve the problem of finding the infrared asymptotics
$s\equiv k/\mu\rightarrow 0$, because the corresponding perturbation theory is a series in
the parameter $s^{-2\eps}$ growing without limit in the region we are interested in. The problem is
solved by passing to the RG representation. To use it for the response functions
(\ref{Gff}) and (\ref{Gtt}), rewrite them in the renormalized variables in the form
\begin{align}
\Gamma_{\varphi\varphi'}(k,\omega = 0) &= \nu k^2 R_{\varphi}(s,g),\nonumber\\
\quad \Gamma_{\psi\psi'}(k,\omega = 0) &= u \nu k^2
R_{\psi}(s,g,u), \label{RG}
\end{align}
where the dimensionless functions $R_{\varphi}$ and $R_{\psi}$ of dimensionless arguments
$s$, $g$, and $u$ are given by the expressions
\begin{eqnarray}
R_{\varphi}(s,g) = Z_\nu - \frac
{\Sigma_{\varphi'\varphi}(k,\omega = 0)}{\nu k^2 }\,,\nonumber\\
R_{\psi}(s,g,u) = Z_1 - \frac {\Sigma_{\psi'\psi}(k,\omega = 0)}{u
\nu k^2}\, .
 \label{RRR}
\end{eqnarray}
The RG representation for functions (\ref{RG}) is determined by the relations
\begin{align}
\Gamma_{\varphi\varphi'}(k,\omega=0) &= \bar \nu k^2
R_{\varphi}(s=1,\bar g)\,, \nonumber \\
 \Gamma_{\psi\psi'}(k,\omega=0)&= \bar u \bar \nu k^2
R_{\psi}(s=1,\bar g, \bar u)\,, \label{RR1}
\end{align}
where $\bar g=\bar g(s,g)$, $\bar \nu=\bar \nu(s,g,\nu)$, and $\bar
u=\bar u(s,g,u)$ are invariant variables satisfying RG equations of the form
\[
 \big[ - s\partial_s + \beta_g\partial_{g}+\beta_u \partial_{u}
-\gamma_{\nu}\,\nu \partial_\nu\big] b(s,g,u) = 0,
 %\label{RGeq}
\]
and normalized by the conditions $\bar g(1,g)=g$, $\bar \nu(1,g,\nu)=\nu$, and
$\bar u(1,g,u)=u$.  The RG functions $\beta$ and
$\gamma$ are defined by the renormalization constants according to the relations
\begin{align}
 \beta_g(g) &\equiv \mu \partial_\mu
 \big |_{0}
g = g(-2\varepsilon+3\gamma_\nu), &\beta_u(g,u)& \equiv \mu
\partial_\mu
 \big |_{0} u = u (\gamma_1 - \gamma_{\nu})\,,\nonumber\\
 \gamma_\nu(g) &\equiv \mu \partial_\mu
 \big |_{0} \ln Z_{\nu}\,,    & \gamma_1(g,u)&\equiv \mu \partial_\mu
 \big |_{0}
  \ln Z_1, \label{RGF1}
\end{align}
where $\mu \partial_\mu
 \big |_{0}$ denotes the operator $ \mu
\partial_{\mu} $ acting at fixed bare parameters
$g_0$, $\nu_0$, and $u_0$. The last equalities for the $\beta$ functions in Eq.
(\ref{RGF1}) are a consequence of the
connections between the renormalization constants in Eq. (\ref{Z}).

As shown in the one-loop approximation in Ref.
\cite{Hnatich,Pismak,Teodorovich88}, the invariant charges $\bar g(s,g)$ and
$\bar u(s,g,u)$ in the limit $s \rightarrow 0$ tend to the
infrared-stable fixed point:\, $\bar g(s,g)\rightarrow
g_*=O(\eps)$,  $\bar u(s,g,u) \rightarrow u_*=O(\eps^0)$, and the
invariant viscosity has the powerlike asymptotic behavior
\[
\bar\nu = \left( {D_{0} k^{-2\eps} \over \bar g }\right)^{1/3}
\to \left( {D_{0} k^{-2\eps} \over g_{*}} \right)^{1/3} \, .
\]
Thus, the expression for the effective inverse Prandtl number (\ref{ueff}) in the inertial range
predicted by the RG representation with the account of relations
(\ref {RG}) and (\ref {RR1}) is
\begin{eqnarray}
u_{eff} =u_* \frac
{R_{\psi}(s=1,\,g_*,\,u_*)}{R_{\varphi}(s=1,\,g_*)} \,.
 \label{ueff1}
\end{eqnarray}

\section{\label{sec:Pr}Two-loop calculation of the Prandtl number}

The expansion of the functions $R_{\varphi}$  and $R_{\psi}$ (\ref{RRR}) in the coupling constant
$g$ is of the form
\begin{align}
R_{\varphi} &= 1 + g \,\left[{a_{11}^{(\nu)}\over\eps} -
A_{\varphi}\,s^{-2\eps}\right] + O(g^2),\nonumber\\
R_{\psi} &= 1 +  g
\,\left[{a_{11}^{(1)}(u)\over\eps }- A_{\psi}(u) s^{-2\eps}\right]+ O(g^2)\,.
 \label{RR}
\end{align}
Here, the quantities $A_{\varphi}$ and $A_{\psi}$ are determined by the one-loop contribution
to $\Sigma_{\varphi'\varphi}$ and $\Sigma_{\psi'\psi}$, whereas the coefficients
$a_{11}^{(\nu)}$ and $ a_{11}^{(1)}$ of representation
(\ref{ZZ}) of the renormalization constants $Z_{\nu}$ and $Z_1$ are found from the condition of
UV finiteness of expressions (\ref{RR}). Substituting relations (\ref{RR})
in Eq. (\ref{ueff1}) we obtain
\begin{align}
u_{eff}  &= u_* \left\{1+[a_{\varphi} - a_{\psi}(u_*)] g_* +
O\left(g^2_*\right)\right\}\,, \label{ueff2}\\
a_{\varphi} &\equiv A_{\varphi}-
{a_{11}^{(\nu)}\over\eps}, \qquad a_{\psi}\equiv A_{\psi}(u_*) -
{a_{11}^{(1)}(u_*)\over\eps}\,.\label{aa}
\end{align}
Bearing in mind that $g_*=O(\eps)$, we see that to find
$u_{eff}$ at the leading order of the $\eps$ expansion, it is enough to
know the charge $u_*$ in the one-loop approximation. At the second order, apart from the
more accurate values of $u_*$ and $g_*$, it is necessary to calculate
the coefficients $a_{\varphi}$ and $a_{\psi}(u_*)$ of the expansion of the scaling functions
(\ref{RRR}) and (\ref{RR}) at the leading order in $\eps$ as well.

The location of the fixed point ($g_*$, $u_*$) is determined by the conditions
$\beta_g(g_*)=\beta_u(g_*,u_*)=0$. The nontrivial fixed point with
$g_*\neq 0$  is infrared stable \cite{Hnatich}, and
from Eq. (\ref{RGF1}) the relations
\begin{align}
\gamma_\nu(g_*)&=\frac{2\eps}{3}\,, \label{gnu}\\
\gamma_1 (g_*,u_*)&=\frac{2\eps}{3}\,\label{g1}
\end{align}
follow at this fixed point.

\begin{figure}
\includegraphics[width=7cm]{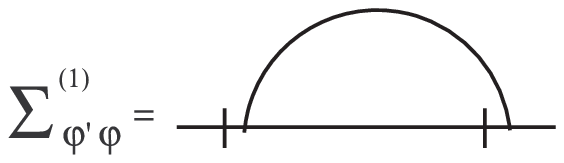}
\caption{\label{fifi} The one-loop self-energy graph for $\Sigma_{\varphi'\varphi}$.
The lines correspond to propagators (\ref{lines0}) and (\ref{lines}). Slashes denote the end carrying
arguments of the field $\varphi '$; plain end carries the arguments of the $\varphi$ field.
Vertices correspond to the factor $V_{ijs}={\rm i}(k_j\delta_{is}+k_s\delta _{ij})$.}
\end{figure}

\begin{figure}
\includegraphics[width=6.4cm]{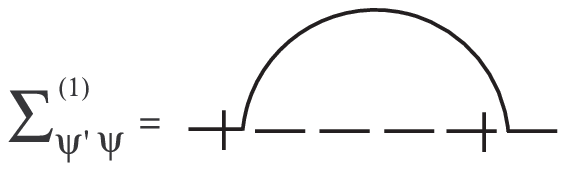}
\caption{\label{psipsi} The one-loop self-energy graph for $\Sigma_{\psi'\psi}$.
The lines correspond to propagators (\ref{lines0}) and (\ref{lines1}). Slashes denote the end carrying
arguments of the field $\psi'$; plain end carries the arguments of the field $\varphi$ or $\psi$.
Vertices correspond to the factor $V_{ijs}={\rm i} k_j$.}
\end{figure}

The UV-finiteness of the RG functions $\gamma(g,u)$ from Eq. (\ref{RGF1}) allows to express them in terms
of the coefficient of the first-order pole in $\eps$ in expression
(\ref{ZZ}) for the renormalization constants:
\begin{equation}
\gamma=(\beta_g\partial_g+\beta_u\partial_u)\ln Z=-2g\partial_g
a_1\,.\label{ga1}
\end{equation}
The renormalization constant $Z_\nu$ at the second order of perturbation theory and the corresponding
expression for $\gamma_\nu$ were obtained in Ref.
\cite{2loop}. For $g_*$ from Eq. (\ref{gnu}) the result is:
\begin{equation}
\label{FPd0}
g_{*} \bar S_{d} = \frac{8(d+2)\eps}{3(d-1)} \,(1+\lambda\eps) + O(\eps^{3}),
\qquad \bar S_{d}\equiv {S_d\over(2\pi)^d}\,,
\end{equation}
where
\begin{equation}
\lambda \simeq
-1.101,\quad d = 3;\qquad \lambda = - \frac{2}{3(d-2)}+ c +
O(d-2),\quad d \rightarrow 2\,.
 \label{FPd}
\end{equation}
From previous analyses
the renormalization constant $Z_1$ is known in the one-loop approximation only \cite{Hnatich}:
\begin{eqnarray}
Z_1 = 1 +\frac{a_{11}^{(1)}\,g}{\eps} +\left( \frac{A}{\eps^2} +
\frac{B}{\eps}\right)\,\left(g\bar S_{d}\right)^2 + O\left(g^3\right)\,,\quad a_{11}^{(1)}
= - \frac{(d-1)\bar S_{d}}{4du(1+u)} \,.\label{Z1}
\end{eqnarray}
Calculation of the contributions $A(u)$ and $B(u)$ of order $g^2$ is presented below
[it should be noted that, like the one-loop factor $a_{11}^{(1)}$, the two-loop coefficients
of the poles in $\eps$ in representation (\ref{Z1}) are non-polynomial functions of $u$].
According to Eq. (\ref{ga1}),
the RG function $\gamma_1$ corresponding to Eq. (\ref{Z1}) is
\[
\gamma_1 = \frac{(d-1)g\bar S_{d}}{2du(1+u)} - 4B(g\bar S_{d})^2
+O(g^3)\,.
%\label{gamma}
\]
Iterative solution of Eq. (\ref{g1}) with respect to $u$ with the account of relation (\ref{FPd0}) yields
\begin{align}
u_* &= u_*^{(0)} + u_*^{(1)}\eps  + O\left(\eps^2\right)\,,\qquad
u_*^{(0)}\left[1+u_*^{(0)}\right]= {2(d+2)\over d}\,,\label{u}\\
u_*^{(1)}&=
\frac{2(d+2)}{d\left[1+2u_*^{(0)}\right]}\left[\lambda-\frac{128(d+2)^2}{3(d-1)^2}B\left(u_*^{(0)}\right)\right]\,.
 \label{u1}
\end{align}
Substituting relations (\ref{u}) and (\ref{u1}) in Eq. (\ref{ueff2}) and taking into account Eq.
(\ref{FPd0}) we obtain
\begin{equation}
u_{eff}=
u^{(0)}_*\left(1+\eps\left\{\frac{1+u^{(0)}_*}{1+2u^{(0)}_*}\,
\left[\lambda-\frac{128(d+2)^2}{3(d-1)^2}B\left(u_*^{(0)}\right)\right] +
\frac{8(d+2)}{3(d-1)\bar S_{d}}(a_{\varphi} - a_{\psi})\right\}\right
)+O\left(\eps^2\right)\,.
 \label{uefff}
\end{equation}
We now turn to the calculation of the constants
$B$, $a_{\psi}$ and $a_{\varphi}$ which determine the Prandtl number. In the one-loop approximation
the quantities
$\Sigma_{\varphi'\varphi}$ and $\Sigma_{\psi'\psi}$ are represented by the graphs
depicted in Figs. \ref{fifi} and \ref{psipsi}, respectively.
In the one-loop self-energy graphs of Figs. \ref{fifi} and \ref{psipsi},
the lines correspond to propagators (\ref{lines0}), (\ref{lines}) and (\ref{lines1})
with the convention that ends with slashes corresponds to
arguments of the fields $\varphi '$ and $\psi'$, plain ends of $\varphi$ and $\psi$. Vertices in Figs. \ref{fifi} and
\ref{psipsi} correspond to the factors $V_{ijs}={\rm i}(k_j\delta
_{is}+k_s\delta _{ij})$ and ${\rm i} k_j$, respectively. Upon contraction of indices,
integration over time and introduction of dimensionless wave vector
(in units of the external wave vector $\p$) in the integrals
we obtain
\begin{align}
A_{\varphi} &= \frac {1}{2(d-1)}\int \frac {d \k} {(2\pi)^d}\,
\frac {k^{2-d-2\eps}(1 - \xi^2)[2 k^3\xi - (d-3)k^2- 2 k(d-1) \xi
-(d-1)] }{(2 k^2 + 2k \xi +1 )(k^2 + 2k \xi +1 )}\, ,
 \label{Afi}\\
A_{\psi}(u) &= - \frac {1}{2u}\int \frac {d \k} {(2\pi)^d}\, \frac
{k^{2-d-2\eps}(1 - \xi^2)}{(1+u)k^2 + 2 u k \xi +u }\, , \qquad \xi
\equiv {\k \p\over(kp)}\,.
 \label{Apsi}
\end{align}
The integrals (\ref{Afi}) and (\ref{Apsi}) are UV divergent in the limit $\eps \rightarrow 0$,
the residue at the pole is readily found by selecting the asymptotic at $k \rightarrow \infty$ contributions
to the integrands and discarding the inessential region of integration $k \leq 1$. Thus, for the coefficients
$ a_{\varphi}$ and  $a_{\psi}$ together with the renormalization constants $Z_{\nu}$ and
$Z_1$ chosen to cancel divergences in expressions (\ref{aa}) we find
\begin{align}
\frac{a_{11}^{(\nu)}}{\eps} &= \frac
{1}{4(d-1)(2\pi)^d}\int\limits_1^{\infty}\frac {dk}{k^{1+2\eps}} \int
d\hat{\k}\, (1 - \xi^2)(2 k\xi-d+3 -6\xi^2)\, ,\nonumber\\
%\label{at}\\
\frac{a_{11}^{(1)}(u)}{\eps} &= \frac
{-1}{2u(1+u){(2\pi)^d}}\int\limits_1^{\infty} \frac
{dk}{k^{1+2\eps}} \int d\hat{\k} \,(1 - \xi^2),  \qquad
\hat{\k}\equiv {\k\over k}, \,.\nonumber
%\label{apsi}
\end{align}
Replacing the integral over directions of the unit vector
$\hat{\k}$ by the average over its directions $\int d\hat{\k}... =
S_d \langle ...\rangle$ and taking into account that
\begin{equation}
\langle \xi^{2n}\rangle=\frac{(2n-1)!!}{d(d+2)\dots(d+2n-2)},
\qquad \langle \xi^{2n+1}\rangle=0,\label{ugly}
\end{equation}
we arrive at result (\ref{Znu}) for $a_{11}^{(\nu)}$ and
(\ref{Z1}) for $a_{11}^{(1)}$. In view of the preceding argumentation, the coefficients $a_{\varphi}$ and
$a_{\psi}$ in Eq. (\ref{aa}) at the leading order in $\eps$ may be written as
\begin{multline}
a_{\varphi} = \frac
{1}{4(d-1)(2\pi)^d}\int\limits_0^{\infty}\! dk \int\! d\hat{\k}\, (1 -
\xi^2)\Biggl \{\frac {2k[2 k^3\xi - (d-3)k^2- 2 k(d-1) \xi -(d-1)]
}{(2 k^2 + 2k \xi +1 )(k^2 + 2k \xi +1 )}\label{afid}\\
-{\theta(k-1)(2 k\xi-d+3 -6\xi^2)\over k}\Biggr \}\,,
\end{multline}
\begin{eqnarray}
  a_{\psi} = \frac
{-1}{2u(2\pi)^d}\int\limits_0^{\infty}\! dk \int\! d\hat{\k}\, (1 -
\xi^2)\left[ \frac {k}{(1+u)k^2 + 2 u k \xi +u }-
\frac{\theta(k-1)}{k(1+u)}\right]\,.
\label{apsid}
\end{eqnarray}
At $d=3$ from relations (\ref{afid}) and (\ref{apsid}) we obtain
\begin{align}
  a_{\varphi} &= \frac
{\bar S_{3}}{8}\int\limits_0^{\infty}\! dk \int\limits_{-1}^1 \!d\xi\, (1 -
\xi^2)\left \{\frac {2k[k^3\xi - 2 k \xi -1] }{(2 k^2 + 2k \xi +1
)(k^2 + 2k \xi +1 )}\label{afi3} -\theta(k-1)\left(\xi
-{3\xi^2\over k}\right)\right \},\\
  a_{\psi} &= \frac
{-\bar S_{3}}{4u}\int\limits_0^{\infty}\! dk \int\limits_{-1}^1 \!d\xi\, (1 -
\xi^2)\left [ \frac {k}{(1+u)k^2 + 2 u k \xi +u }-
\frac{\theta(k-1)}{k(1+u)}\right]\label{apsi3},\quad d=3,\,\,\,\bar S_{3}=\frac{1}{2\pi^2}\,,
\end{align}
Numerical evaluation of integrals (\ref{afi3}) and (\ref{apsi3}) with
$u=u_*^{(0)}$ from Eq. (\ref{u}) yields
\begin{equation}
a_{\varphi}= -0.047718\,\bar S_{3}\,,\quad a_{\psi}=-0.04139\,\bar
S_{3}\,.
\label{afipsi}
\end{equation}

It is convenient to find
the two-loop contributions to the renormalization constant
$Z_1$ from the condition that the quantity $R_{\psi}$ from Eq. (\ref{RRR}) is UV finite in the limit
$k\rightarrow 0$. In terms of the reduced quantity
\begin{equation}
\Sigma \equiv \lim_{k\rightarrow 0} \frac
{\Sigma_{\psi',\psi}(\omega=0,k)}{u \nu k^2}
 \label{sigma}
\end{equation}
this condition may be cast in the form
\begin{equation}
Z_1(\eps) - \Sigma(\eps) = O(\eps^0).
 \label{sss}
\end{equation}
The limit $k\rightarrow 0$ in expression (\ref{sigma}) does exist, provided the IR regularization
of the graphs has been taken care of. In the MS scheme renormalization constants do not depend on the
method of such regularization. With our choice of the pumping function
(\ref{nakach}) it is accomplished by the cutoff of the propagator $\langle \varphi \varphi \rangle _0$
(\ref{lines0}) at $k < m$.

Let us choose further the wave vector of integration such that in the lines
$\langle \varphi \varphi \rangle _0$ it flows alone (for the graphs $\Sigma_{\psi',\psi}$ such a choice is
always possible). Then integration over all the wave numbers will be carried out within the
limits from $m$ to $\infty$.

The one-loop contribution to $\Sigma$ is determined by the graph of Fig.
\ref{psipsi} as:
\begin{multline}
\Sigma^{(1)} = - \frac {g\,\mu^{2\eps}}{2u(Z_\nu+uZ_1)Z_{\nu}}\int\!
\frac {d \k} {(2\pi)^d}\, \frac {(1 -
\xi^2)\theta(k-m)}{k^{d+2\eps}}
\\  = - \frac
{g\,\mu^{2\eps}}{2u(Z_\nu+uZ_1)Z_{\nu}(2\pi)^d}\int\limits_m^{\infty}\! \frac
{dk}{k^{d+2\eps}}\int d\hat{\k} \,(1 - \xi^2) = - \frac
{g\,\bar S_{d}\,\mu^{2\eps}}{2u(Z_\nu+uZ_1)Z_{\nu}}\int\limits_m^{\infty}\!
\frac {dk}{k^{d+2\eps}} \,\langle(1 - \xi^2)\rangle\,,
%\label{s1}
\nonumber
\end{multline}
which, together with relations (\ref{Znu}), (\ref{Z1}), and (\ref{ugly}) yields
\begin{equation}
\Sigma^{(1)} = - \frac {g\,\bar S_{d}\,(d-1)\tau^{-2\eps}}{4\,
\eps \,u\,(Z_\nu+uZ_1)\,Z_{\nu}\, d}= - \frac {g\,\bar
S_{d}\,(d-1)\tau^{-2\eps}}{4\, \eps \,u\,(1+u)\,\, d}\,\left\{1 -
\left[u\,a_{11}^{(1)}+(2+u)a_{11}^{(\nu)}\right]\,\frac{ g \,\bar S_{d}}{\eps(1+u)}\,\right\}+O(g^3)
\label{ss1},
\end{equation}
where $\tau \equiv m/\mu$. Extracting the pole contributions in $\eps$ from
expressions (\ref{ss1}) we obtain
\begin{equation}
\Sigma^{(1)} = - \frac {g\,\bar S_{d}\,(d-1)}{4\, \eps
\,u\,(1+u)\,\, d}\,\left\{1 -
\left[u\,a_{11}^{(1)}+(2+u)a_{11}^{(\nu)}\right]\,\frac{g \,\bar S_{d}
}{(1+u)}\left(\frac{1}{\eps} - 2 \ln \tau\right)\,\right\}+O(\eps^0)
\label{ss2}.
\end{equation}
Substituting relation (\ref{ss2}) in Eq. (\ref{sss}) and requiring cancellation of pole contributions
in the linear in $g$ approximation, we return to expression
(\ref{Z1}) for $a_{11}^{(1)}$. The terms of order $g^2$ are required for the calculation of the renormalization
constant in the two-loop approximation.

\begin{figure}
\includegraphics[width=13cm]{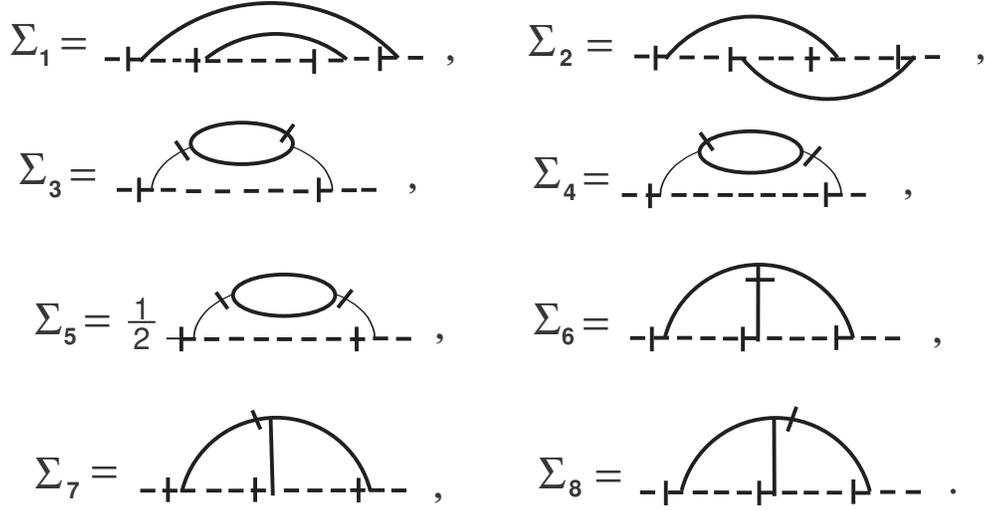}
\caption{\label{fig} The two-loop self-energy graphs for $\Sigma_{\psi'\psi}$.
The lines correspond to propagators (\ref{lines0}), (\ref{lines}) and (\ref{lines1}). Slashes denote the end carrying
arguments of the field $\varphi'$ or $\psi'$; plain end carries the arguments of the field $\varphi$ or $\psi$.
Vertices correspond to the factor  $V_{ijs}={\rm i}(k_j\delta_{is}+k_s\delta _{ij})$ or $V_{ijs}={\rm i} k_j$.}
\end{figure}

The two-loop contribution $\Sigma^{(2)}$ to the self-energy operator
$\Sigma_{\psi'\psi}$ is determined by the sum of the graphs depicted in Fig. \ref{fig}
[normalization according to Eq. (\ref{sigma}) is implied].
When substituting propagators (\ref{lines0}), (\ref{lines}) and (\ref{lines1}) -- expressed in terms of the renormalized
variables -- in the graphs of
Fig. \ref{fig} it is possible to put $Z_{\nu}=Z_1=1$ with the necessary accuracy. Contracting indices and
integrating over time we obtain
\begin{eqnarray}
  \Sigma_n = \frac {(g \bar S_{d})^2 \mu^{4\eps}}{192 u v^2}\int\limits_m^\infty \!\frac {d k}{k^{1+2\eps}}
  \int\limits_m^\infty\! \frac{d q}{q^{1+2\eps}}
   \int\limits_{-1}^1\! d \xi\, \frac{(1-\xi^2)\,J_n}{[ v
 (k^2+q^2)+u
k q \xi]} \,,\quad n = 1,\,2\,,
 \label{sigma1,2}
\end{eqnarray}
where
\begin{eqnarray}
 J_1 = 2 q^2\,,\qquad  J_2 = - z k q \,,
 \label{12u}
\end{eqnarray}
and
\begin{eqnarray}
  \Sigma_n = \frac {(g \bar S_{d})^2 \mu^{4\eps}}{96 u v}\int\limits_m^\infty \!
  \frac {d k}{k^{1+2\eps}} \int\limits_m^\infty\! \frac{d q}{q^{1+2\eps}}
   \int\limits_{-1}^1\! d \xi\, \,\frac{(1-\xi^2)q^2\, J_n}{k^2+2kq\xi+q^2} \,,\qquad n = 3\ldots 8\,,
 \label{sigma3..8}
\end{eqnarray}
with
\begin{align}
 J_3 &=  { k (k^3+2 k^2 q \xi -q^3 \xi)}
\, \left[\frac{1}{(k^2+kq\xi+q^2)(v k^2+ k q \xi+q^2)}+
\frac{1}{v k^2(v k^2+ k q \xi+q^2)}+\frac{1}{k^2( k^2+ k q
\xi+q^2)}\right]\,, \label{3u}\\
 J_4 &= \frac { (k^3+2 k^2 q \xi -q^3 \xi)}{ k( k^2+
k q \xi+q^2)}
 \label{4u}\,,\\
 J_5 &= -\,\frac {k^2[k^4+q^4+k q \xi (k^2+q^2)]}{(k^2+q^2+k q \xi)
 (k^2+q^2+ 2 k q \xi)}\, \left[\frac {2}{k^2+2 k q \xi+q^2}+
 \frac{1}{v(k^2+q^2)+ u k q \xi}\right]\,,
 \label{5u}\\
 J_6 &= \frac {k q \xi (k^2 - q^2)}{2(k^2+ k q \xi +v q^2)}\,\left(\frac{1}{v
 q^2}+\frac{1}{k^2+kq\xi+ q^2}\right)
  \label{6u}\,,\\
 J_7 &= -\, \frac { k^3 (2 k^3+3 k^2 q \xi -q^3 \xi)}{2 ( k^2+
2 k q \xi+q^2)}\,\Biggr\{ \frac{1}{v k^2 [v(k^2+q^2)+u k q \xi ]}\nonumber\\
&\phantom{\frac {(2 k^3+3 k^2 q \xi -q^3 \xi)}{2 ( k^2+
2 k q \xi+q^2)}\Biggr\{ \frac{1}{ [v(k^2+q^2)+u k q \xi ]}}+
\frac{1}{v k^2 (v k^2 +k q \xi +q^2)}+\frac{1}{(k^2+k q \xi
+q^2)(v k^2 + k q \xi +q^2)} \Biggr\}\,,
 \label{7u}\\
 J_8 &= \, \frac { k (2 k^3+3 k^2 q \xi -q^3 \xi)}{2(k^2 +k q \xi +q^2)[v(k^2+q^2)+ukq\xi]}
 \label{8u}\,.
\end{align}
Integrals (\ref{sigma1,2}) - (\ref{8u}) may be represented as
\begin{equation}
\Sigma_{i}(\eps)=\mu^{4\eps}\int\limits_{m}^{\infty}\!
\frac{dk}{k^{1+2\eps}} \int\limits_{m}^{\infty}\! \frac{dq}{q^{1+2\eps}}
\int\limits_{-1}^{1}\!d\xi\, f_{i}(\xi,k/q), \label{si}
\end{equation}
or, after the corresponding stretching of integration variables, as
\begin{equation}
\Sigma_{i}(\eps)=\int\limits_{\tau}^{\infty} \!\frac{dk}{k^{1+2\eps}}
\int\limits_{\tau}^{\infty} \!\frac{dq}{q^{1+2\eps}} \int\limits_{-1}^{1}\!d\xi\,
f_{i}(\xi,k/q),\qquad \tau \equiv m/\mu \,,\label{si1}
\end{equation}
or, finally, as
\begin {equation}
\Sigma_{i}(\eps)=A_i \tau^{-4\eps}\,, \qquad A_i(\eps) \equiv
\int\limits_{1}^{\infty} \!\frac{dk}{k^{1+2\eps}} \int\limits_{1}^{\infty}\!
\frac{dq}{q^{1+2\eps}} \int\limits_{-1}^{1}\!d\xi \,f_{i}(\xi,k/q) .
\label{si2}
\end{equation}
We are interested in the coefficients of the pole contributions to
$\Sigma_{i}(\eps)$:
\begin{equation}
A_i=\frac{a_{i}}{\eps^{2}}+\frac{b_{i}}{\eps}+O(\eps^0)\,,\quad
\Sigma_{i}(\eps)=\frac{a_{i}}{\eps^{2}}+\frac{b_{i}-4 a_i \ln \tau
}{\eps}+O(\eps^0). \label{A20}
\end{equation}
For the functions $f_{i}(z,k/q)$ with $i=2,\, 5 ... 8$ the equations
$f_{i}(z,0)=f_{i}(z,\infty)=0$ hold revealing that integrals over
$k$ and $q$ in Eq. (\ref{si}) are separately convergent, so that the divergence at
$\eps\to0$ in the corresponding $\Sigma_i$ is brought about by the region, in which
$k$ and $q$ tend to infinity simultaneously. As a consequence, the second-order pole is
absent in such $\Sigma_i$: $a_{i}=0$ for $i=2,\, 5 ... 8$.

For $\Sigma_{i}$ with $i=1,3,4$  $f_{i}(z,\infty)=0$ as before, which means absence of divergence
in the integral over $k$ in Eq. (\ref{si}). For these graphs, however,
$f_{i}(z,0)={\rm const}\ne0$, so that the integral over $q$ diverges at $\eps \rightarrow 0$ leading
to the appearance of the pole of second order in the full integral.

Expressions
(\ref{si1}) may be simplified with the use of the identity
\begin{equation}
\Sigma_{i}=-\frac{\tau \partial_{\tau}\Sigma_{i}}{4\eps}
\label{mpartial}
\end{equation}
following from Eq. (\ref{si2}). Calculating the right-hand side of Eq. (\ref{mpartial})
with the aid of relations (\ref{si1}) and introducing the dimensionless integration variables,
we obtain
\begin{equation}
\Sigma_{i}(\eps)=\frac{\tau^{-4\eps}}{4\eps}\int\limits_{1}^{\infty}\!
\frac{d\kappa}{\kappa^{1+2\eps}} \int\limits_{-1}^{1}\!d\xi \left[
f_{i}(\xi,\kappa)+f_{i}(\xi,1/\kappa) \right]. \label{A21}
\end{equation}
This operation has reduced the number of iterated integrations and allowed for explicit extraction
one pole in $\eps$. For $i=2,\,5...8$
the integral in Eq. (\ref{A21}) is finite for $\eps=0$ and determines the residue of the first-order pole:
\begin{equation}
a_{i}=0,\qquad b_{i}=\frac{1}{4}\int\limits_{1}^{\infty}\!
\frac{d\kappa}{\kappa} \int\limits_{-1}^{1}\!d\xi
\left[f_{i}(\xi,\kappa)+f_{i}(\xi,1/\kappa)\right], \quad
i=2,\,5...8. \label{A22}
\end{equation}
For $\Sigma_{i}(\eps)$ with $i=1,3,4$ the coefficient of the second-order pole is obtained
by the replacement of the function
$\left[f_{i}(\xi,\kappa)+f_{i}(\xi,1/\kappa)\right]$ in the integrand in Eq. (\ref{A21})
by its limiting value at $\kappa\rightarrow \infty$:
$f_{i}(\xi,\infty)+f_{i}(\xi,0)=f_{i}(\xi,0)$ [we remind that
$f_{i}(z,\infty)=0$]. Then integration over $\kappa$ becomes trivial, which yields
\begin{equation}
a_{i}=\frac{1}{8}\int\limits_{-1}^{1}\!d\xi f_{i}(\xi,0), \quad i=1,3,4.
\label{A23}
\end{equation}
The remaining integral with the change
$f_{i}(\xi,\kappa)\to\left[f_{i}(\xi,\kappa)-f_{i}(\xi,0)\right]$ is finite at $\eps=0$
and determines the residue of the first-order pole:
\begin{equation}
b_{i}=\frac{1}{4}\int\limits_{1}^{\infty}\! \frac{d\kappa}{\kappa}
\int\limits_{-1}^{1}\!dz
\left[f_{i}(z,\kappa)+f_{i}(z,1/\kappa)-f_{i}(z,0)\right], \quad
i=1,3,4. \label{A24}
\end{equation}
Let us write condition (\ref{sss}) at order $g^2$ for $d=3$. With the use of the corresponding
terms of the one-loop contribution (\ref{ss2}), the summed two-loop contributions
(\ref{A20}) and expression
(\ref{Z1}) for the renormalization constant $Z_1$, we obtain
\begin{eqnarray}
 \frac{A}{\eps^2}+\frac{B}{\eps} = \frac {1}{6\, \eps
\,u\,(1+u)^2}\,
\left[u\,a_{11}^{(1)}+(2+u)a_{11}^{(\nu)}\right]\,\left(\frac{1}{\eps} - 2 \ln
\tau\right)+\sum_1^8\left(\frac{a_{i}}{\eps^{2}}+\frac{b_{i}-4 a_i \ln
\tau }{\eps}\right)\,.
 \label{finit}
\end{eqnarray}
With the aid of expressions (\ref{sigma1,2})--(\ref{4u}) and (\ref{si}) in Eq. (\ref{A23}), it is not difficult to find
\begin{equation}
a_1 =\frac{(g \bar S_{3})^2}{72u(1+u)^3}, \quad a_3 =\frac{(g \bar
S_{3})^2(3+u)}{480u(1+u)^2}, \quad a_4 =\frac{(g \bar
S_{3})^2}{480u(1+u)},\quad d=3.
%\label{a134}
\nonumber
\end{equation}
Substituting these values in Eq.
(\ref{finit}) and taking into account relations
(\ref{Z1}) and (\ref{Znu}) for $a_{11}^{(1)}$ and $a_{11}^{(\nu)}$,
we see that the terms with $\ln \tau$ in Eq. (\ref{finit}) are automatically cancelled
(as a consequence of renormalizability of the model), whereas for the coefficient $A$ of the
second-order pole we obtain
\begin{equation}
A =- \,\frac {3u^2+9u+16}{720u(1+u)^3}\,.
%\label{A}
\nonumber
\end{equation}

\begin{table}
\caption{\label{tab:table1}
Residues of the first-order poles in $\eps$ of the dimensionless integrals
(\ref{si2}) corresponding to the two-loop graphs of Fig. \ref{fig}.
}
\begin{ruledtabular}
\begin{tabular}{lcccccccc}
$ i $& 1 & 2 & 3 & 4 & 5 & 6 & 7 &8\\
\hline
$b_i\cdot 10^3$& $0.1099$ & $0.0944$ & $0.8691$ & $0.0057$& $-3.9382$ &$0.0672$ & $-1.9647$ &$0.5899$\\
\end{tabular}
\end{ruledtabular}
\end{table}
For the coefficients
$b_i$ numerical integration of expressions
(\ref{A22}) and (\ref{A24}) with $u=u_*^{(0)}$ from Eq. (\ref{u}) yields
the results quoted in Table \ref{tab:table1},
which for the coefficient $B$ in Eq. (\ref{finit}) lead to the value
\begin{equation}
B\left( u_*^{(0)}\right) =\sum_{i=1}^8 b_i = - 4.1666 \cdot 10^{-3}\,.
%\label{B}
\nonumber
\end{equation}
Substituting this value in Eq. (\ref{uefff}) as well as $a_{\varphi}$ and
$a_{\psi}$ from Eq. (\ref{afipsi}) and $\lambda$ from Eq. (\ref{FPd}), we obtain the final
expression for the effective inverse Prandtl number:
\begin{equation}
u_{eff}= u^{(0)}_*(1-0.0358\eps)+O(\eps^2)\,,\qquad
u^{(0)}_*={\sqrt{43/3}-1\over 2}\simeq 1.3930\,, \qquad d=3\,.
\label{otvet}
\end{equation}
At the physical value $\varepsilon=2$ this yields for the turbulent Prandtl number Pr$_t$
the result:
\begin{equation}
\label{PrT}
\text{Pr}_t^{(0)}\simeq 0.7179 \,,\qquad \text{Pr}_t\simeq 0.7693
\end{equation}
in one-loop and two-loop accuracy, respectively.

\section{\label{sec:Zakl}Conclusion}

The main conclusion to be drawn from the two-loop value of the effective inverse Prandtl number
(\ref{otvet}) obtained in the present paper is that the correction term is strikingly small. Even at the real
value $\eps=2$ it is only 7\% of the leading contribution. Apparently this is the reason of the
favorable comparison of the one-loop value of the
turbulent Prandtl number 0.72 \cite{Fournier,Hnatich} with the experiment: the recent circular-jet result
$0.81\pm 0.05$ \cite{Chua90} is corroborated by the experimentally recommended value 0.8 for modeling
and once more confirmed the range 0.7 -- 0.9 of measured values \cite{Chang02} which has been put forward, however,
quite a while ago \cite{Monin}. In view of these numbers it may be concluded that the already fairly good one-loop result
is improved by the two-loop correction, whose account
(\ref{PrT}) leads to the value 0.77  for the turbulent Prandtl number.
At the same time this result is somewhat unexpected -- similar two-loop corrections
to the Kolmogorov constant and the skewness factor are large \cite{2loop}.

In the results obtained there are, however, also significant features common with the calculation
\cite{2loop}. From Table \ref{tab:table1} of coefficients
$b_i$ we see that $b_5$ has the largest value while the value of the whole sum
$B\left( u_*^{(0)}\right)= \sum_{i=1}^8 b_i$ is close to that of $b_5$ alone. The graph $\Sigma_5$, which gives rise to this
coefficient, is the only one of the two-loop graphs of Fig.
\ref{fig} possessing a singularity in space dimension $d=2$. Exactly the same situation was encountered
also in Ref. \cite{2loop} in the two-loop calculation of the constant
$\lambda\simeq -1.101$  in Eq. (\ref{FPd}). It is a rather unexpected observation that for terms
in the factor
\,$M=\lambda-800B/3\simeq 0.010$ from Eq. (\ref{uefff}) ($d=3$) an almost complete numerical cancellation
takes place. To clarify the situation, we calculated asymptotics of the graph $\Sigma_5$
at $d\rightarrow 2$ with the result:  $b_5 \simeq -1/1024(d-2)$. Substitution of this expression
together with the analogous one for $\lambda$ from Eq. (\ref{FPd}) in the factor
$M(d)$ reveals that the singular in $d-2$ contributions indeed cancel, so that this factor and with it
the whole two-loop contribution to the Prandtl number are finite at $d=2$!
It should be noted that also in the second term of the two-loop contribution to $u_{eff}$ [see Eq. (\ref{uefff})]
a significant decrease in magnitude takes place in the difference  $a_{\varphi} - a_{\psi}$
of the contributions from renormalization of viscosity
and thermal diffusivity compared with the magnitudes of these terms separately.

Thus, our results complement the conclusion made in Ref. \cite{2loop}. In the two-loop approximation the main
contribution is due to graphs having a singularity at
$d=2$ and it is necessary to sum such graphs. For quantities in which this singularity is absent the two-loop
contribution is relatively small.

\end{document}